\documentclass{aa}

\usepackage{graphicx}

\usepackage{txfonts}
\usepackage[normalem]{ulem}
\usepackage{color, colortbl}
\usepackage{dirtytalk}
\usepackage [english]{babel}
\usepackage [autostyle, english = american]{csquotes}
\MakeOuterQuote{"}
\usepackage{amsmath}
\usepackage{hyperref}
\usepackage{soul}

\begin{document}

   \title{X-ray polarization observations of NGC 2110 with { IXPE}}

   \author{I. Pal
          \inst{1}, S. Marchesi \inst{1, 2, 3}, N. Torres-Albà \inst{1,4}, 
          I. Cox \inst{1}, M. Ajello \inst{1}, A. Banerjee \inst{1}, R. Silver \inst{5}, A. Pizzetti\thanks{ESO Fellow} \inst{1,6} \and K. Imam \inst{1}
          }

   \institute{Department of Physics and Astronomy, Clemson University, Kinard Lab of Physics, Clemson, SC 29634, USA \\
              \email{ipal@clemson.edu}
         \and
             Dipartimento di Fisica e Astronomia (DIFA), Università di Bologna, via Gobetti 93/2, I-40129 Bologna, Italy
             \and
                INAF - Osservatorio di Astrofisica e Scienza dello Spazio di Bologna, Via Piero Gobetti, 93/3, 40129, Bologna, Italy
                \and
                    Department of Astronomy, University of Virginia, P.O. Box 400325, Charlottesville, VA 22904, USA
                    \and
                        NASA Goddard Space Flight Center, Greenbelt, MD 20771, USA
                    \and
                        European Southern Observatory, Alonso de Córdova 3107, Casilla 19, Santiago 19001, Chile.
                        }

   \date{Received September 15, 1996; accepted March 16, 1997}

   \abstract
   {X-ray polarimetric observations from the Imaging X-ray Polarimeter Explorer ({IXPE}) is an excellent tool for probing the geometry and dynamics of X-ray emitting corona in active galactic nuclei (AGNs).}
   {This work aims to investigate the geometry of the X-ray corona in the Seyfert 2 AGN, NGC 2110, using its first polarimetric observation with { IXPE}, conducted over a net exposure of 554 ks beginning on October 16, 2024.}
   {We performed a model-independent analysis of the 2–8 keV { IXPE} polarimetric observation to estimate the polarization properties of NGC 2110. Furthermore, we performed spectral and spectro-polarimetric analyses combining { IXPE} data with archival observations from {NuSTAR}, {XMM-Newton}, and {Swift-XRT} to derive detailed spectral and polarization properties.}
   {From the spectro-polarimetric analyses, an upper limit on the polarization degree ($\Pi_{X}$) of 7.6$\%$ (at the 99$\%$ confidence) was estimated in the 2$-$8 keV band. The spectro-polarimetric analysis in the 5.66$-$8 keV band yielded a looser upper limit of $\lesssim$27\% at the 99$\%$ confidence. This result aligns closely with the findings from the model-independent analysis.}
   {Comparing the measured polarization properties, coronal parameters, and inclination angle of NGC 2110 with the Monte Carlo radiative transfer (MONK) simulations suggest that the current polarization measurements lack the sensitivity to place definitive constraints on the coronal geometry. The upper limits on $\Pi_{X}$, as derived from our analysis at the 99\% confidence level, indicate that polarization remains undetected at a statistically significant level. Consequently, we are unable to determine whether the corona is elongated along the disk or more compact and spherical. Future observations with improved sensitivity will be crucial to breaking these degeneracies and providing deeper insight into the coronal structure of NGC 2110.}

    
\authorrunning{Pal et al.} 

\maketitle

\section{Introduction}
Active galactic nuclei (AGNs), among the most luminous objects in the Universe, are powered by the accretion of matter onto the super-massive black holes (SMBHs) at the center of galaxies \citep{1964ApJ...140..796S, 1969Natur.223..690L, 1973A&A....24..337S, 1983MNRAS.205..593G, 1984ARA&A..22..471R, 2008ARA&A..46..475H}. AGNs emit across the entire electromagnetic spectrum and are widely recognized for the intense X-ray emission \citep{1978MNRAS.183..129E, 1993ARA&A..31..717M}. The primary X-ray emission in Seyfert type AGNs is believed to originate from inverse Compton scattering of the optical-UV seed photons by the hot plasma ($\sim$$10^{8-9}$K), commonly known as corona situated near the accretion disk \citep{1975ApJ...195L.101T, 1980A&A....86..121S, 1991ApJ...380L..51H, 2000ApJ...542..703Z, 2007A&ARv..15....1D}. Recent studies, most notably using observations from the Nuclear Spectroscopic Telescope Array ({NuSTAR}, \citealt{2013ApJ...770..103H}), were able to characterize the properties of the corona in radio-quiet AGNs (e.g., its temperature, optical depth, etc, \citealt{refId0,2019MNRAS.484.2735M,2020ApJ...905...41B,2022ApJ...927...42K}). However, despite numerous studies in the literature aimed at characterizing the corona, its geometry remains poorly understood, potentially holding vital clues to its physical origin. The launch of {IXPE} \citep{10.1117/1.JATIS.8.2.026002} has enabled the revolutionary studies of the X-ray polarization from radio-quiet AGN coronae for the very first time, since X-ray polarimetry can independently probe coronal properties and the geometry of the corona \citep{2022MNRAS.516.5907M, 10.1093/mnras/stad2627, 2023JApA...44...87P, 2023MNRAS.525.5437I, 2023MNRAS.523.4468G, 2024A&A...691A..29G}.

Over the years, several morphological models have been proposed for the X-ray emitting corona. At least four distinct configurations have been widely suggested to describe the corona in AGNs: 1) slab \citep{1991ApJ...380L..51H, 2017ApJ...850..141B}, 2) wedge-shaped \citep{1997ApJ...489..865E, 1996ApJ...470..249P, 2014ARA&A..52..529Y, 2018A&A...614A..79P, 2020A&A...634A..92U}, 3) spherical lamppost \citep{1991A&A...247...25M, 2012MNRAS.424.1284W, 10.1093/mnras/stab3745}, and 4) conical outflow \citep{1997A&A...326...87H, 2004A&A...413..535G, 10.1093/mnras/stab3745}. The launch of {IXPE} enables us to measure the polarimetric properties of radio-quiet AGNs, opening a unique window to place constraints on these proposed models. These configurations have different polarization signatures. In the slab corona model, the hot medium is uniformly distributed above the cold accretion disk, originating from magnetic instabilities. The slab corona can generate a $\Pi_{X}$ of up to 14\%, with the polarization angle ($\Psi_{X}$) parallel to the accretion disk axis \citep{1996ApJ...470..249P, 10.1093/mnras/stab3745, 2023MNRAS.523.4468G}. The wedge-shaped hot accretion flow resembles the slab corona with a height that increases with radius. In this scenario, the standard accretion disk is truncated at a certain radius, and the corona represents a hot accretion flow, possibly extending down to the innermost stable circular orbit (ISCO). This configuration is expected to produce an intermediate $\Pi_{X}$ (up to 5\%, depending on the inner, outer radius and the opening angle), with $\Psi_{X}$ parallel to the accretion disk~axis \citep{10.1093/mnras/stad2627}. Thus, when comparing the slab and wedge-shaped coronae, the former produces a higher $\Pi_{X}$, while the latter results in an intermediate $\Pi_{X}$. In both cases,  $\Psi_{X}$ remains parallel to the accretion disk.

The spherical lamppost model features an isotropic spherical structure located along the spin axis of the supermassive black hole. This structure is predicted to generate a low $\Pi_{X}$ (0-2\%) with $\Psi_{X}$ perpendicular to the accretion disk axis \citep{10.1093/mnras/stab3745}. Finally, the conical outflow model is often linked to an aborted jet scenario. This configuration is expected to produce a slightly higher $\Pi_{X}$ (up to 6\%), with the $\Psi_{X}$ also perpendicular to the accretion disk axis \citep{10.1093/mnras/stab3745}. Therefore, when combined with $\Pi_{X}$, $\Psi_{X}$ can help us break the degeneracy between different coronal geometries. The simulations using the { MONK} code \citep{Zhang_2019}, \cite{10.1093/mnras/stab3745} demonstrated that the $\Pi_{X}$ for a given geometry can vary with the spectral shape, coronal temperature, optical depth, the inclination of the system and the black-hole spin. However, for spherical and conical geometries, $\Pi_{X}$ is always significantly lower than for the slab geometry, with a $90^\circ$ difference in $\Psi_{X}$ between the slab/wedge and the spherical lamppost/conical outflowing coronae.

NGC 2110 is a nearby Seyfert galaxy at a redshift of $z$ = 0.007849. It has a black hole mass of $\sim$ 2 $\times$ $10^{8}$ M$_{\odot}$ \citep{2007ApJ...668L..31M}. NGC 2110 is an early-type spiral galaxy (SAB0) with a Seyfert 2 nucleus, located in an S0/E host galaxy at a distance of 30.4 Mpc \citep{1985ApJ...291..627W, 1979ApJ...233..809M}. The galaxy is observed at an intermediate inclination angle, estimated between 42$^{\circ}$ and 65$^{\circ}$ \citep{1985ApJ...289..124W, 2002ApJ...579..188G, 2020ApJ...895..135K}. Initially classified as a narrow-line X-ray galaxy, NGC 2110 has a dusty nucleus that obscures the broad-line region, leading to its Seyfert 2 optical spectrum. However, the gas column density is insufficient to attenuate the 2–10 keV emission, resulting in hard X-ray luminosities comparable to Seyfert 1 galaxies \citep{1999A&A...342L..41M}. NGC 2110 is a low-accreting AGN with an Eddington luminosity ($L_{Edd}$) = 2.4 × $10^{46}$ erg $s^{-1}$ and the bolometric luminosity ranges between ($L_{bol}$) = 0.6–9×$10^{44}$ erg $s^{-1}$ \citep{2015MNRAS.447..160M}. One notable characteristic of NGC 2110 is the lack of the reprocessed emission from distant Compton-thick scattering \citep{1999A&A...342L..41M, 2014ApJ...786..126R, 2015MNRAS.447..160M}. The strong iron emission lines observed in the source are believed to originate from Compton-thin material, as the lack of a significant Compton reflection component ($R \lesssim$ 0.2) above 10 keV suggests the reprocessing material is not Compton-thick \citep{1991A&A...247...25M, 1991MNRAS.249..352G}, makes NGC 2110 an excellent candidate for studying coronal geometry through X-ray polarization observations in the 2–8 keV band, where coronal emission dominates. In this work, we present our results on the analysis of the X-ray polarimetric observations carried out on NGC 2110 by { IXPE}. The observations and data reduction are described in Section 2, the analysis and results are described in Section 3,  followed by the summary in Section 4.

\begin{figure*}
\hbox{
     \includegraphics[scale=0.45]{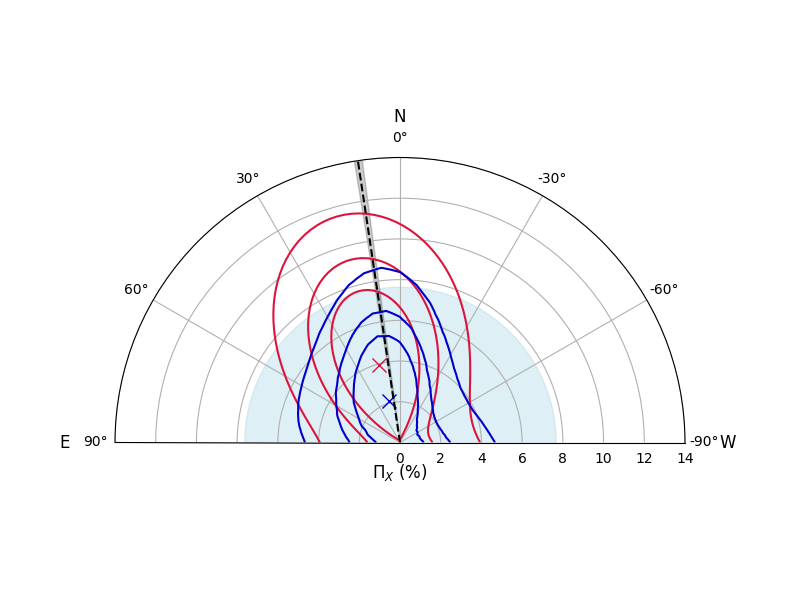}
     \includegraphics[scale=0.45]{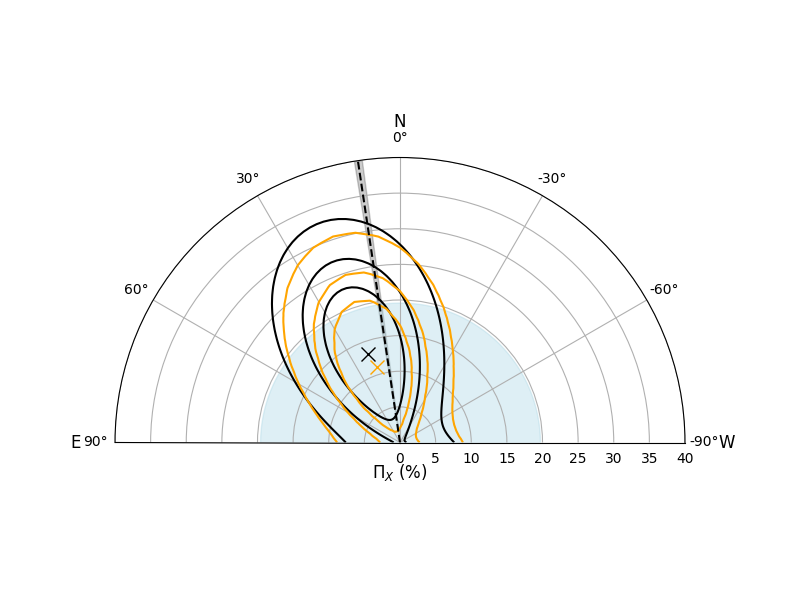}
     }
\caption{Left panel: The contour plots between $\Pi_{X}$ and $\Psi_{X}$ measured in the 2-8 keV band using the $ PCUBE$ (red) and $ XSPEC$ (blue) analyses. The red and blue crosses are the best-fitted values of $\Pi_{X}$ and  $\Psi_{X}$. The contours are plotted in 68\% ($\chi^{2}$ = 2.28), 90\% ($\chi^{2}$ = 4.605) and 99\% ($\chi^{2}$ = 9.21) confidence levels.  Right panel: The polarization degree $\Pi_{X}$ and polarization angle $\Psi_{X}$ contour plots as obtained from the 5.66-8 keV $ PCUBE$ (black) and $ XSPEC$ (orange) analyses. The black and orange crosses are the best-fitted values of $\Pi_{X}$ and  $\Psi_{X}$. The contours are plotted in 68\% ($\chi^{2}$ = 2.28), 90\% ($\chi^{2}$ = 4.605) and 99\% ($\chi^{2}$ = 9.21) confidence. The black dashed line represents the radio jet position angle ($\sim$8.5$^{\circ}$) of NGC 2110. The black shaded region is the error in jet position angle. The blue shaded regions denote the $MDP_{99}$ value at the 99\% confidence.} 
\label{figure-2}
\end{figure*}

\begin{table}
        \centering
        \caption{Observation logs of { IXPE, NuSTAR, XMM-Newton \rm{\&} Swift} observations of NGC 2110.}
        \label{table-1}
\begin{tabular}{cccc}
\hline
Telescope & OBSID  & Date & Exposure Time \\
    &  &       &        (ks)         \\
\hline
{ IXPE} & 03008799 & 2024-10-16 & 	554  \\
        \hline
{ NuSTAR} & 60061061002 & 2012-10-05 & 	16  \\
& 60061061004 & 2013-02-14 & 	12 \\ 
\hline
{ XMM-Newton} & 0145670101 & 2003-03-05 & 60 \\
\hline
{ Swift-XRT} & 35459002 & 2006-04-08 & 8.6 \\
& 35459004 & 2008-08-31 & 2.2 \\
& 35459005 & 2009-10-12 & 3.0 \\
& 80364001 & 2012-10-05 & 6.5 \\
& 35459006 & 2013-02-03 & 13.0 \\
& 92804007 & 2017-10-13 & 4.5 \\
& 35459014 & 2019-08-07 & 3.3 \\
& 35459031 & 2020-04-08 & 3.7 \\
\hline
        \end{tabular}
\end{table}

\section{Observations and data reduction}
Over the past few years, NGC 2110 has been observed by several X-ray observatories. In our study, we used archival data from {IXPE, NuSTAR, XMM-Newton}, and {Swift-XRT}. The log of these observations is provided in Table \ref{table-1}. The data reduction process is detailed in the following subsections.
\subsection{{ IXPE}} 
{ IXPE}  started observing NGC 2110 on October 16, 2024, with its three detector units (DUs) for a net exposure of about 554 ks. The calibrated data were produced by the standard { IXPE} pipeline provided by the Science Operation Center (SOC)\footnote{\href{https://heasarc.gsfc.nasa.gov/docs/ixpe/analysis/IXPE-SOC-DOC-009-UserGuide-Software.pdf}{https://heasarc.gsfc.nasa.gov/docs/ixpe/analysis/IXPE-SOC-DOC-009-UserGuide-Software.pdf}}. We used the cleaned and calibrated level 2 data for the scientific analysis. The cleaned level 2 event files were first treated with the background
rejection procedure described in \cite{2023AJ....165..143D} \footnote{\href{https://github.com/aledimarco/IXPE-background}{https://github.com/aledimarco/IXPE-background}}. Then they were analyzed using the { IXPEOBSSIM} software v30.0.0 \citep{BALDINI2022101194}. A count map in sky coordinates was generated using the { CMAP} algorithm within the { xpbin} task. We adopted a circular region with a radius of $60''$ for the extraction of the source of the three DUs and an annulus with inner and outer radii of $150''$ and $300''$, respectively, for the background extraction of each DU \citep{2023AJ....165..143D}. We then used the { xpselect} task to generate the filtered source and background regions.

For the spectro-polarimetric analysis, the { I, Q, U} source and background spectra were generated using the { PHA1}, { PHA1Q} and { PHA1U} algorithm using { xpbin} task within { IXPEOBSSIM} for the three DUs. The { I, Q, U}  spectra are rebinned using { ftgrouppha} task, where { I} spectra are binned with a minimum of 30 counts/bin, and Stokes spectra are subsequently binned using { I} spectrum as a template file.

\subsection{{ NuSTAR}}
NGC 2110 has been observed twice by {NuSTAR} in 2012 and 2013. Both observations are publicly available and analyzed in this work. We reduced the { NuSTAR} data in the 3$-$79 keV band using the standard { NuSTAR} data reduction software NuSTARDAS\footnote{https://heasarc.gsfc.nasa.gov/docs/nustar/analysis/nustar swguide.pdf} v2.1.2 distributed by HEASARC within HEASoft v6.33.  The calibrated, cleaned, and screened event files were generated by running { nupipeline} task using the CALDB release 20220510. To extract the source counts we chose a circular region of radius $100''$ centred on the source. Similarly, to extract the background counts,  we selected a circular region of the same radius away from the source on the same chip to avoid contamination from source photons. We then used the  { nuproducts} task to generate energy spectra, response matrix files (RMFs) and auxiliary response files (ARFs), for both the hard X-ray detectors housed inside the corresponding focal plane modules, FPMA and FPMB.

\subsection{{ XMM-Newton}}
{ XMM-Newton} observed NGC 2110 once in 2003 with the EPIC CCD cameras: the pn \citep{2001A&A...365L..18S} and the two MOS \citep{2001A&A...365L..27T}. We carried out our analysis with the data from the pn camera. We used SAS v1.3 for the data reduction. The event files were filtered to exclude background flares selected from time ranges where the 10–15 keV count rates in the PN camera exceeded 0.7 c/s. Source spectra were extracted from a circular region with a radius of $50''$ centred on the nucleus. Background photons were selected from a source-free region of equal area on the same chip as the source. We checked for pileup using the { EPATPLOT} task. We did not find the source suffered from the pileup.  We constructed RMFs and ARFs using the tasks { RMFGEN} and { ARFGEN} for each observation. Data from { NuSTAR} and { XMM-Newton} are binned with 30 counts/bin to incorporate $\chi^2$ statistics during fitting. 

\begin{figure}
\hbox{
     \includegraphics[scale=0.38]{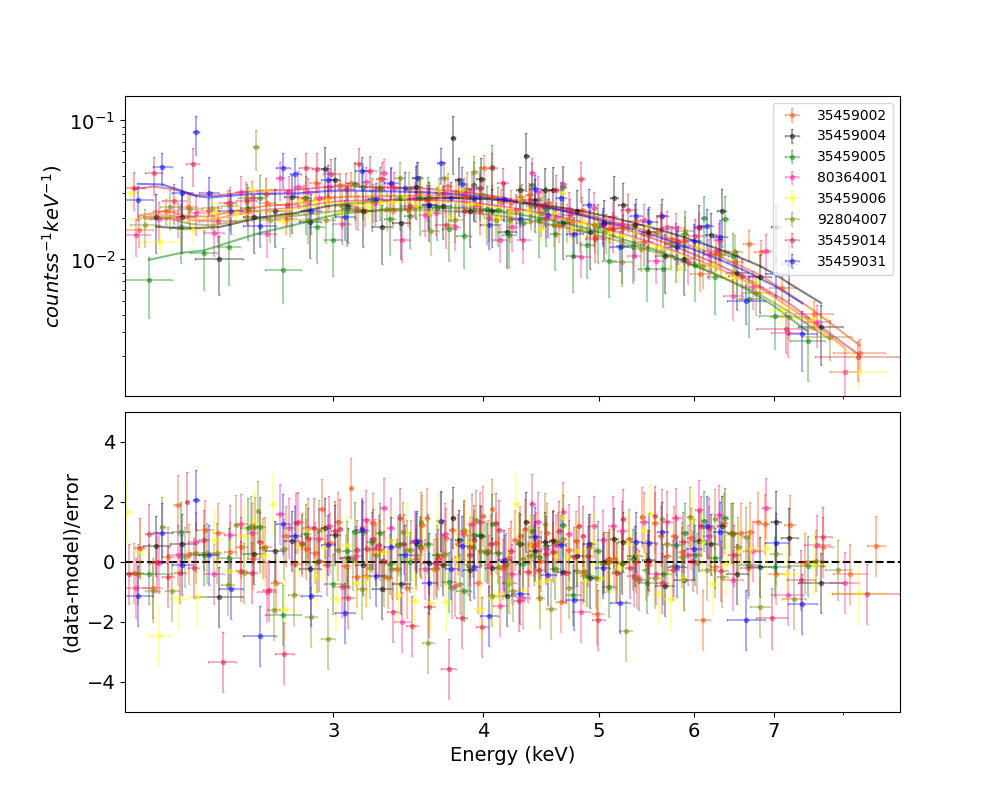}
     }
\caption{{ Swift-XRT} best-fitted spectra (top panel) with $ phabs*zphabs*(zpo)$ and the residuals (bottom panel) in the 2-9 keV band showing no significant spectral variation over years of observations.}
\label{figure-3}
\end{figure}

\subsection{{ Swift-XRT}}
There are 42 { Swift/XRT} observations of NGC 2110, from 2006 to 2020. To check for spectral variation over years, we have selected the observation from each year with maximum exposure (see Table \ref{table-1}). The corresponding spectrum was generated in the 0.5–9 keV energy band using the standard online tools provided by the UK Swift Science Data Centre \citep{2009MNRAS.397.1177E}.

\bigskip

The uncertainties in the spectral parameters derived from the XSPEC analysis correspond to a confidence level 90\% ($\Delta \chi^{2} = 2.706$), unless otherwise stated. The errors associated with the polarization parameters ($\Pi_{X}$ and $\Psi_{X}$) obtained from the polarimetric and spectro-polarimetric analyses are quoted at a 68\% confidence level, while the upper limits on the MDP and polarization parameters are reported at a 99\% confidence level.

\begin{table}
        \centering
        \caption{Polarization parameters in different energy bands as obtained from $ PCUBE$ analysis.}
        \label{table-2}
\begin{tabular}{|c|cccc|}
\hline
Energy & $\Pi_{X}\pm \delta\Pi_{X}$ & $\Psi_{X}\pm \delta\Psi_{X}$   & $\rm{MDP_{99}}$ & $\Pi_{X_{99}}$ \\
& \% & deg. & \% & \% \\
\hline
2-8 &  3.93$\pm$2.52 & 15$\pm$18 & 7.65 & $<$10.48 \\
\hline
2-4 &  2.11$\pm$2.38 & -23$\pm$32 & 8.03 & $<$8.30 \\
4-8 &  7.19$\pm$3.75 & 22$\pm$15 & 11.39 & $<$16.94 \\
\hline
2-2.83 &  1.13$\pm$4.26 & -69$\pm$108 & 11.92 & $<$12.21\\
2.83-4 &  3.77$\pm$2.88 & -17$\pm$22 & 8.72 & $<$11.26 \\
4-5.66 &  1.87$\pm$3.50 & 38$\pm$54 & 10.61 & $<$10.97 \\
5.66-8 &  13.18$\pm$6.47 & 20$\pm$14 & 19.64 & $<$30.00 \\
\hline
\end{tabular}
\tablefoot{The errors in $\Pi_{X}$ and $\Psi_{X}$ are quoted at the 68\% confidence. $\rm{MDP_{99}}$ quantifies the polarization sensitivity at the 99\% confidence level. The upper limits at the 99\% confidence level for one parameter of interest ($\Pi_{X_{99}}$) are also reported.}
\end{table}

\section {Data analysis}
In the following subsections, we demonstrate polarimetric and spectropolarimetric analyses of the {IXPE, NuSTAR, XMM-Newton}, and {Swift-XRT} observations.

\subsection{Polarimetry}
The polarimetric signal of NGC 2110 was analyzed in a model-independent way using the { PCUBE} algorithm based on \cite{2015APh....68...45K} method in the { xpbin} task with the software { ixpeobssim} v30.0.0 \citep{BALDINI2022101194}. The three polarization cubes for the three DUs were generated to extract information such as Stokes parameters; $\rm{MDP_{99}}$; $\Pi_{X}$; $\Psi_{X}$ and their associated errors. We first generated the three polarization cubes corresponding to the three DUs across the entire 2$-$8 keV energy band. The combined polarization parameters are presented in Table \ref{table-2}. From the { PCUBE} analysis, we obtained $\Pi_{X}$ = 3.93\% $\pm$ 2.52\% at the 68\% confidence level, with an upper limit of $\lesssim$ 10.48\% at 99\% confidence. The contours of $\Pi_{X}$ versus $\Psi_{X}$ at the 68\% ($\chi^2$ = 2.28), 90\% ($\chi^2$ = 4.605), and 99\% ($\chi^2$ = 9.21) confidence levels are shown in the left panel of Fig. \ref{figure-2}.

To investigate the energy dependence of the polarization parameters, we derived these parameters in multiple energy bins using the { PCUBE} algorithm. The 2–8 keV range was divided into 2–12 bins (e.g., for two bins: 2–4 keV and 4–8 keV). Table \ref{table-2} presents the polarization parameters for the 2–8 keV data, divided into two and four energy bins. The largest $\Pi_{X}$ value was measured in the 5.66–8 keV band, where we obtained $\Pi_{X}$ = 13.18\% $\pm$ 6.47\% at the 68\% confidence. Contour plots of $\Pi_{X}$ versus $\Psi_{X}$ were generated to evaluate the detection significance in this band (see right panel of Fig. \ref{figure-2}). We obtained an upper limit of 30\% for $\Pi_{X}$. However, considering the entire 2–8 keV band or dividing it into multiple bins, no significant polarization detection can be claimed in any energy range as the measured $\Pi_{X}$ values remain lower than the MDP values. The absence of a significant polarization degree suggests that the polarization angle remains unconstrained.

\begin{figure*}
\hbox{
     \includegraphics[scale=0.38]{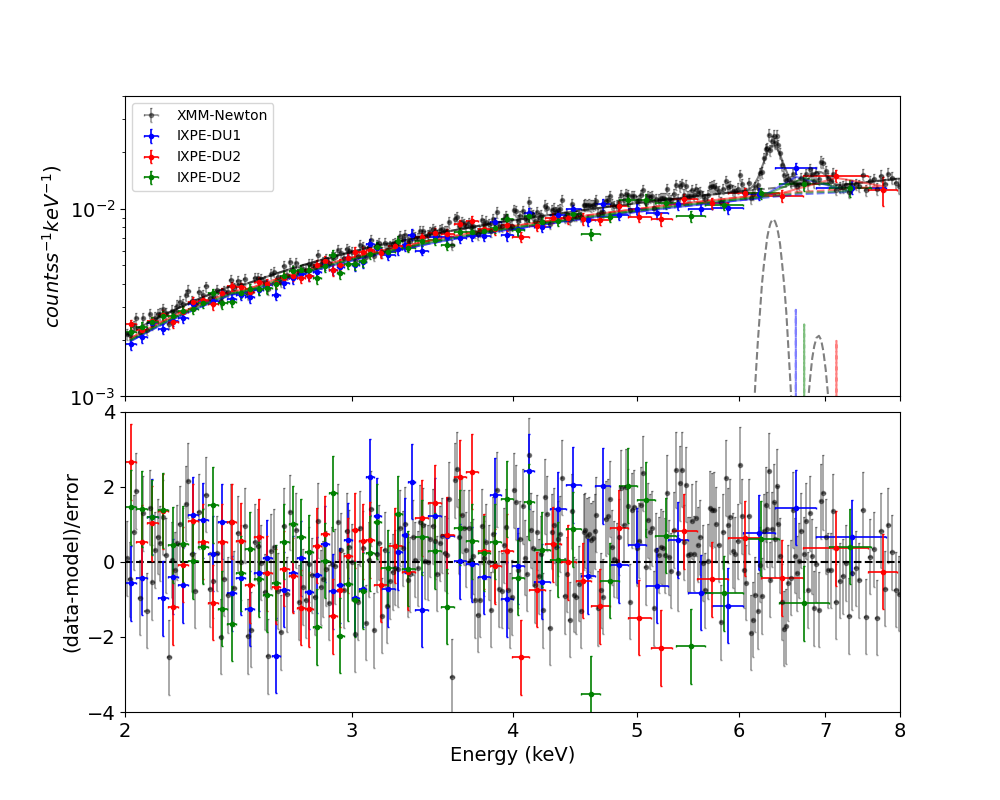}
     \includegraphics[scale=0.38]{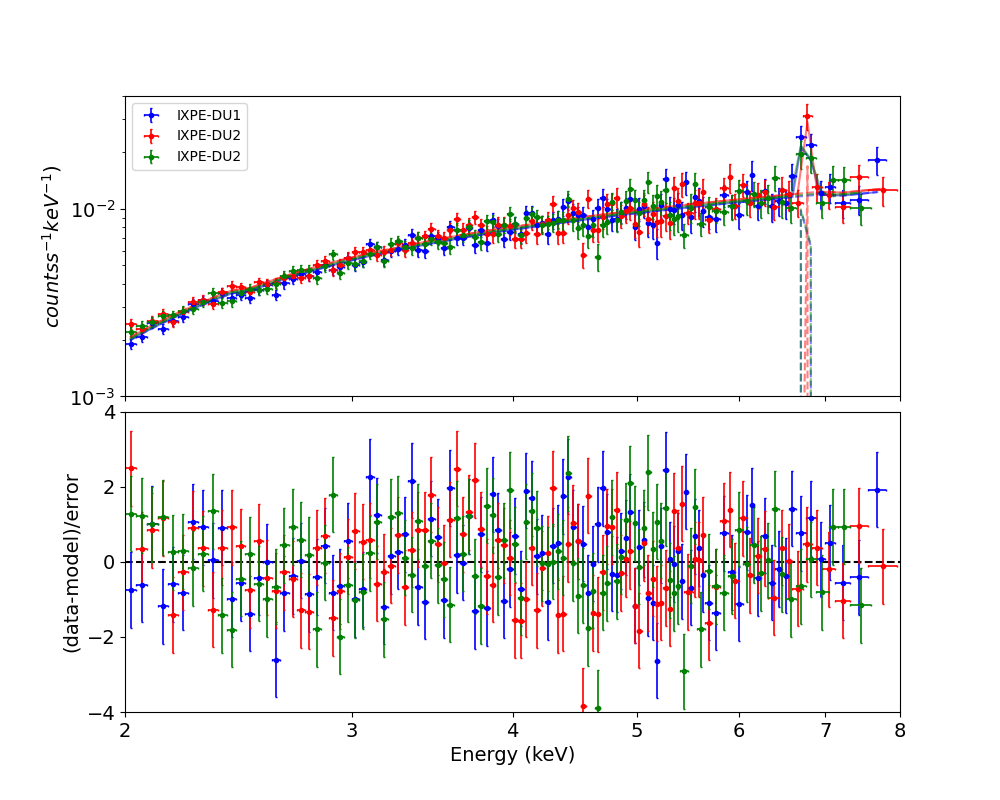}
     }
\caption{Left panel: Best-fitted { XMM-Newton} and { IXPE I-spectra} and right panel: Best-fitted { IXPE I-spectra} with power-law model $ constant*phabs*zphabs*(zpo+zgauss)$ with data to model residues.} 
\label{figure-4}
\end{figure*}

\begin{table}
        \centering
        \caption{Parameters obtained from the best-fitted model $ phabs*zphabs*(zpo)$ to the ${XRT}$ observations of NGC 2110}
        \label{table-4}
\begin{tabular}{ccccc}
\hline
OBSID & $\rm{N_{H,los}}$ & $\Gamma$ & norm & $c_{stat}/\rm{dof}$ \\
& $\times$ $10^{22} cm^{-2}$ & & & \\
\hline
35459002 & 3.63$^{+1.37}_{-1.33}$ & 1.38$^{+0.36}_{-0.35}$ & 0.02$^{+0.02}_{-0.01}$ & 152/167   \\
35459004 & $<$6.45 & $<$1.44 & $<$0.02 & 27/42 \\
35459005 & 6.15$^{+3.15}_{-2.91}$ & 1.56$^{+0.79}_{-0.75}$ & $<$0.15 & 35/50 \\
80364001 & 3.27$^{+1.67}_{-1.60}$ & 1.21$^{+0.45}_{-0.44}$ & 0.03$^{+0.04}_{-0.02}$ & 115/116 \\
35459006 & 2.98$^{+1.15}_{-1.12}$ & 1.14$^{+0.31}_{-0.30}$ & 0.03$^{+0.02}_{-0.01}$ & 221/209 \\
92804007 & 2.57$^{+1.99}_{-1.91}$ & 1.09$^{+0.54}_{-0.52}$ & 0.006$^{+0.010}_{-0.004}$ &  81/87 \\
35459014 & 2.39$^{+1.92}_{-1.84}$ & 1.32$^{+0.54}_{-0.53}$ & 0.007$^{+0.011}_{-0.004}$ & 94/77 \\
35459031 & $<$4.82 & 1.05$^{+0.58}_{-0.47}$ & 0.005$^{+0.008}_{-0.003}$ & 89/79 \\
\hline
\end{tabular}
\end{table}

\subsection{{IXPE, NuSTAR, XMM-Newton} and {Swift-XRT} spectral and spectro-polarimetric analysis}
The X-ray spectra of NGC 2110 have been widely studied in the past few years \citep{2006AN....327.1079R, 2007ApJ...671.1345E, 2015MNRAS.447..160M, 2016ApJS..225...14K, 2020ApJ...905...41B, 2020ApJ...897....2T, 2022ApJ...929..141K, 2023A&A...669A.114D, 2023MNRAS.524.4670J}. Previous studies indicate the presence of intense iron lines in the source spectrum. However, the spectral fitting supports the presence of a very weak Compton reflection with a reflection fraction ($R$) of $\lesssim$ 0.15 \citep{1999A&A...342L..41M, 2014ApJ...786..126R, 2015MNRAS.447..160M, 2023MNRAS.524.4670J}, originating from circumnuclear material partially covered by several layers of absorbing material with column densities in the range 2–6×$10^{22} \rm{cm^{-2}}$ \citep{2015MNRAS.447..160M}. The reprocessed emission from NGC 2110 has been studied using both the reflection and the borus model; however, all models are consistent with a Compton-thin absorber in the line of sight to the primary coronal emission with $\rm{N_{H_{los}}}$ $\sim$ $10^{22}$ atoms $\rm{cm^{-2}}$. Therefore, the previous analyses cumulatively suggest that NGC 2110 is one of the few Seyfert 2 sources where the strong iron lines are not produced by distant, Compton-thick material but are instead emitted by Compton-thin matter. Past analyses performed using several instruments such as, { BeppoSAX, NuSTAR, XMM-Newton, Swift-XRT \& Suzaku} have shown that the spectra of NGC 2110 can be well fitted with an absorbed power-law continuum with photon index ($\Gamma$) $\sim$ 1.7. This suggests that the source has not undergone any dramatic spectral shape alteration over the years of observations. However, there is evidence of a change in the intensity of the iron K lines with the flux state of the source on time scales of years \citep{2015MNRAS.447..160M}. Therefore, we left the line components free to vary during the joint spectral fit. The previous studies on the broadband analysis of NGC 2110 spectra \citep{2006AN....327.1079R, 2007ApJ...671.1345E} also reported a presence of the extra-nuclear emission in the softer-energy (E$<$1 keV) part of the spectra. Since the analysis of that component is beyond the scope of this work, our analysis was limited to the  E$\gtrsim$2 keV band.

\subsubsection{Spectral analysis with an absorbed power-law}
To investigate any spectral variations over the years, we performed the spectral fitting of the eight { Swift-XRT} observations taken between 2006 and 2020. Due to the limited signal-to-noise ratio (S/N), the 2$-$9 keV {XRT} data were binned with a minimum of 5 counts per bin, and the $c$-statistics were applied during the fitting process. Each { XRT} spectrum was fitted independently using a simple absorbed power-law model: ${phabs*zphabs*(zpo)}$. In this model, $phabs$ accounts for the Milky-Way hydrogen column density, which was fixed at 3.01$\times$$10^{22}$ atoms $\rm{cm^{-2}}$, as adopted from \cite{2013MNRAS.431..394W}. The best-fit parameters, presented in Table \ref{table-4}, indicate no significant variation in $\Gamma$ or $\rm{N_{H_{los}}}$ within the uncertainties, therefore, suggesting no prominent spectral evolution of NGC 2110. We continued the absorbed power-law fitting with the 2012 and 2013 { NuSTAR} observations separately in an energy band that partially coincides with the { IXPE} band (i.e., 3-8 keV). The Fe K$\alpha$ line emission in the { NuSTAR} spectrum was modeled with a $gaussian$ component. The 3$-$8 keV spectral analysis of the 2012 { NuSTAR} data with the model $ constant*phabs*zphabs*(zpo+zgauss)$ yielded $\Gamma$ = 1.65$\pm$0.05, $\rm{N_{H_{los}}}$ = 3.96$\pm$0.43 $\times$$10^{22}$ atoms $\rm{cm^{-2}}$, energy of the Fe K$\alpha$ line ($Line_{E}$) = 6.26$\pm$0.07 keV, the calibration constant between FPMA and FPMB ($C_{FPMA/FPMB}$) = 1.01$\pm$0.01, $\chi^{2}/dof$ = 243/242 and 3-8 keV flux ($f_{3-8}$) = 1.50$\times 10^{-10}$ erg $cm^{-2}$ $s^{-1}$. A similar analysis of the 2013 { NuSTAR} data produced $\Gamma$ = 1.67$\pm$0.07, $\rm{N_{H_{los}}}$ = 5.35$\pm$0.96 $\times$$10^{22}$ atoms $\rm{cm^{-2}}$, $Line_{E}$ = 6.42$\pm$0.06 keV, $C_{FPMA/FPMB}$ = 1.00$\pm$0.01, $\chi^{2}/dof$ = 239/242  and $f_{3-8}$ = 1.01$\times 10^{-10}$ erg $cm^{-2}$ $s^{-1}$. The Fe K$\alpha$ line widths were frozen at 100 eV for both { NuSTAR} observations as allowing them to vary did not improve the fit statistics. Fitting the 2$-$8 keV { XMM-Newton PN} data with the absorbed power-law model produced similar best-fit parameters. To take care of the Fe K$\beta$ line component in { XMM} spectrum we had to add another $zgauss$ component and thus, the new model looks like, $ phabs*zphabs*(zpo+zgauss+zgauss)$. The spectral parameters obtained are: $\Gamma$ = 1.45$\pm$0.04, $\rm{N_{H_{los}}}$ = 4.17$\pm$0.24 $\times$$10^{22}$ atoms $\rm{cm^{-2}}$, energy of the Fe K$\alpha$ line ($Line_{E1}$) = 6.42$\pm$0.01 keV, line width ($\sigma_{1}$) = 0.05$\pm$0.02 keV, energy of the Fe K$\beta$ line ($Line_{E1}$) = 6.91$^{+0.14}_{-0.21}$ keV, $\sigma_{2} <$ 0.28 kev,  $\chi^{2}/dof$ = 1061/1094  and $f_{3-8}$ = 1.71$\times 10^{-11}$ erg $cm^{-2}$ $s^{-1}$.

Next, we included the { IXPE} { I}-spectra in our analysis. Although NGC 2110 showed no significant spectral shape variations over years of observation, we detected flux variations between the 2003 { XMM-Newton} and the 2012–2013 { NuSTAR} observations in the 3–8 keV band, which partially overlaps with the { IXPE} range. Therefore, we fitted the { IXPE} data separately with the { NuSTAR} and { XMM-Newton} observations. The Fe line components for the { NuSTAR} and { XMM-Newton} observations were fixed to the best-fit values obtained earlier, and the model parameters were tied across instruments. The calibration constants between the instruments were varied. The results from the absorbed power-law fit are summarized in Table \ref{table-6}. The calibration constants between observations indicate that the { IXPE} 2024 observation ($f_{2-8}$ $\simeq$ 1.74$\times 10^{-11}$ erg $cm^{-2}$ $s^{-1}$) corresponds to a flux state similar to the 2003 { XMM-Newton} observation ($f_{2-8}$ = 1.97$\times 10^{-11}$ erg $cm^{-2}$ $s^{-1}$), as further supported by the best-fit spectra and residuals shown in Fig. \ref{figure-4}.

Since the spectral parameters of the joint fit with { IXPE}, { NuSTAR}, and { XMM-Newton} are primarily driven by the higher S/N ratio of the latter two, we also performed an independent absorbed power-law fit to the 2–8 keV { IXPE} { I}-spectra. The best-fit parameters were: $\Gamma$ = 1.49$\pm$0.10, $\rm{N_{H_{los}}}$ = 4.08$\pm$0.43 $\times$$10^{22}$ atoms $\rm{cm^{-2}}$, energy of the Fe K$\alpha$ line ($Line_{E1}$) = 6.81$\pm$0.30 keV, $\chi^{2}/dof$ = 344/320  and $f_{2-8}$ = 1.74$\times 10^{-11}$ erg $cm^{-2}$ $s^{-1}$. The width of the iron line was frozen to 10 eV.

\subsubsection{Spectro-polarimetric analysis}
For the spectro-polarimetric fit, we simultaneously fitted the { IXPE} { I}-spectra and Stokes spectra ({IXPE Q} and {U}-spectra) from all three DUs with an absorbed power-law ($zphabs*zpo$) convolved with the polarization model $polconst$. A $zgauss$ component was added to model the weak Fe emission line in the { IXPE I-spectra}. Since iron lines are expected to be unpolarized \citep{2011MNRAS.415.3119G, 2018MNRAS.478..950M}, we fixed the polarization of the $zgauss$ components to zero. The final model was: $ constant*phabs*zphabs*(polconst*zpo+polconst^{0}*zgauss)$, where $polconst^{0}$ represents zero polarization. The $constant$ account for the calibration differences between the DUs, with DU1 fixed at 1.0, and DU2 and DU3 yielding 1.00$\pm$0.13 and 0.99$\pm$0.14, respectively.

From the spectro-polarimetric fit in the 2-8 keV band, we obtained $\Pi_{X}$ = 2.12$\pm$2.05\% and $\Psi_{X}$ = 14$^{\circ}$$^{+43^{\circ}}_{-31^{\circ}}$ at the 68\% confidence, with an upper limit of $\Pi_{X}$ $<$7.6\% at the 99\% confidence. The contours $\Pi_{X}$ versus $\Psi_{X}$ are shown in Fig. \ref{figure-2}. Since the highest $\Pi_{X}$ was measured in the 5.66-8 keV band of the algorithm $ PCUBE$, we performed a spectropolarimetric analysis in this range, obtaining $\Pi_{X}$ = 11.16$\pm$6.32\% and $\Psi_{X}$ = $18^{\circ}\pm17^{\circ}$ at 68\% confidence, with an upper limit of $\Pi_{X}$ $<$ 27\% at 99\% confidence. The corresponding contours $\Pi_{X}$ versus $\Psi_{X}$ are also shown in Fig. \ref{figure-2}. The best-fitted { IXPE Q,} { U-} spectra are plotted in Fig. \ref{figure-1}.

\begin{table}
        \centering
        \caption{Best-fitted parameters obtained using the model $constant*phabs*zphabs*(zpo+zgauss)$ to the joint {IXPE} and {NuSTAR}/{XMM-Newton} observations of NGC 2110.} 
        \label{table-6}
\begin{tabular}{lccc}
\hline
\small{Parameter} & ${\small{IXPE}}$ & ${\small{IXPE}}$  & ${\small{IXPE}}$   \\
& +\footnotesize{NuSTAR (2012)} & +\footnotesize{NuSTAR (2013)} & +\footnotesize{XMM}\\
\hline
$\rm{N_{H,los}}$ & 4.70$^{+0.20}_{-0.20}$ & 4.56$^{+0.22}_{-0.22}$ & 4.51$^{+0.13}_{-0.13}$ \\
$\Gamma$ & 1.65$^{+0.03}_{-0.03}$ & 1.61$^{+0.04}_{-0.04}$ & 1.45$^{+0.04}_{-0.04}$ \\
norm & 6.2$^{+0.3}_{-0.3}$ & 4.0$^{+0.3}_{-0.3}$ & 0.51$^{+0.03}_{-0.03}$ \\
$\chi^{2}/\rm{dof}$ & 613/564 & 608/564 & 1447/1418 \\
$C_{FPMA/FPMB}$ & 1.015$^{+0.009}_{-0.010}$ & 1.000$^{+0.014}_{-0.014}$ & - \\
$C_{FPMA/DU1}$ & 0.096$^{+0.002}_{-0.002}$  & 0.141$^{+0.003}_{-0.003}$ & - \\
$C_{FPMA/DU2}$ & 0.100$^{+0.002}_{-0.002}$ & 0.147$^{+0.003}_{-0.003}$ & - \\
$C_{FPMA/DU3}$ & 0.098$^{+0.002}_{-0.002}$ & 0.144$^{+0.003}_{-0.003}$ & - \\
$C_{XMM/DU1}$ & - & - & 0.90$^{+0.01}_{-0.01}$ \\
$C_{XMM/DU2}$ & - & - & 0.93$^{+0.01}_{-0.01}$\\
$C_{XMM/DU3}$ & - & - & 0.91$^{+0.01}_{-0.01}$ \\
\hline
\end{tabular}
\tablefoot{$\rm{N_{H,los}}$ and the normalizations are in units of $10^{22}$ $\rm{atoms}$ $\rm{cm}^{-2}$ and $10^{-2}$.}
\end{table}

\subsubsection{Coronal properties of NGC 2110 with { NuSTAR}}
To begin examining the coronal properties of NGC 2110, we fitted the 3$-$78 keV 2012 { NuSTAR} (OBSID - 60061061002) observation with the following model:
\begin{equation}
    constant \times phabs \times (zphabs \times {zcutoff} + pexrav + zgauss)
\end{equation}

The primary emission was modeled using an absorbed cutoff-power law, { zphabs$\times$zcutoff}. For the reprocessed emission, we used the reflection model $ pexrav$ with a fixed inclination angle of 60$^{\circ}$ \citep{2015MNRAS.447..160M}.  The iron K$\alpha$ emission line in the {NuSTAR} spectrum was modeled with a $zgauss$ component. The best-fit parameters from the spectral analysis of the 2012 { NuSTAR} spectrum are as follows: $\Gamma$ = 1.69$^{+0.01}_{-0.04}$, cut-off energy ($\rm{E_{cut}}$) = 484.50$^{+u}_{-210.22}$ keV \footnote{$u$ in the error denotes an upper limit on the parameter}, reflection-fraction ($R$) $<$ 0.13. The $\chi^{2}$/dof for the fit was 1392/1407. During fitting, $\Gamma$, $\rm{E_{cut}}$ and the model normalizations were tied between ${zcutoff}$ and $pexrav$.  Next, we proceed by adding the 2013 { NuSTAR} observation to the best-fitted model and performed an joint analysis by leaving the model parameters and nrmalizations vary between two observations. For the newly added observation (OBSID - 60061061004), we obtained, $\Gamma$ = 1.70$^{+0.04}_{-0.05}$, $\rm{E_{cut}}$ = 317.67$^{+u}_{-137.49}$ keV and $R$ $<$ 0.15. the line widths were frozen at 10 eV  for both observations. Our analysis agrees well with the findings of \cite{2015MNRAS.447..160M}.  

Since, the joint analysis of the two { NuSTAR} observations did not show any changes in the spectral properties and the source was found to be at a similar flux state, we proceed with the joint analysis by tying all the spectral parameters together. We used the physically motivated comptonization model $xillverCP$ \citep{2014ApJ...782...76G} to calculate the coronal temperature ($\rm{kT_{e}}$) of the source. From the joint fit, we obtained the line-of-sight column density ($N_{H}$) = 5.99$\pm$0.23 $\times$ $10^{22}$ atoms $\rm{cm^{-2}}$, $\Gamma$ = 1.74$\pm$0.01 and $\rm{kT_{e}}$ $>$ 74.86 keV. The $\chi^{2}/dof$ of the fit was 1414/1406. The Fe-K$\alpha$ line was self-consistently taken care of by $xillverCP$. We also varied the inclination angle ($i$) during the spectral fit and obtained a lower limit of $i$ $>$ 74$^{\circ}$. The upper limit of the optical depth ($\tau$) as calculated using the following equation \citep{1996MNRAS.283..193Z,
1999MNRAS.309..561Z},

\begin{equation}
\label{equ2}
    \tau = \sqrt{\frac{9}{4} + \frac{3}{\theta\Big[\Big(\Gamma + \frac{1}{2}\Big)^2 - \frac{9}{4}\Big]}} - \frac{3}{2}  \\
\end{equation} 
is $<$ 1.61, where $\theta = {kT_e}/{m_{e}c^2}$.

Next, we performed a joint analysis of the { NuSTAR}, { XMM-Newton}, and { Swift-XRT} observations using our best-fit model:

\begin{equation*}
\text{constant} \times \text{phabs} \times ( \text{zphabs} \times {zcutoff} + \text{pexrav} + \text{zgauss} )
\end{equation*}

in the $\gtrsim$3 keV band. For this joint analysis, all 42 { Swift-XRT} observations taken between 2006 and 2020 were combined and grouped with a minimum of 50 counts per bin. 

\begin{figure*}
\hbox{
\hspace{0.2cm}
     \includegraphics[scale=0.35]{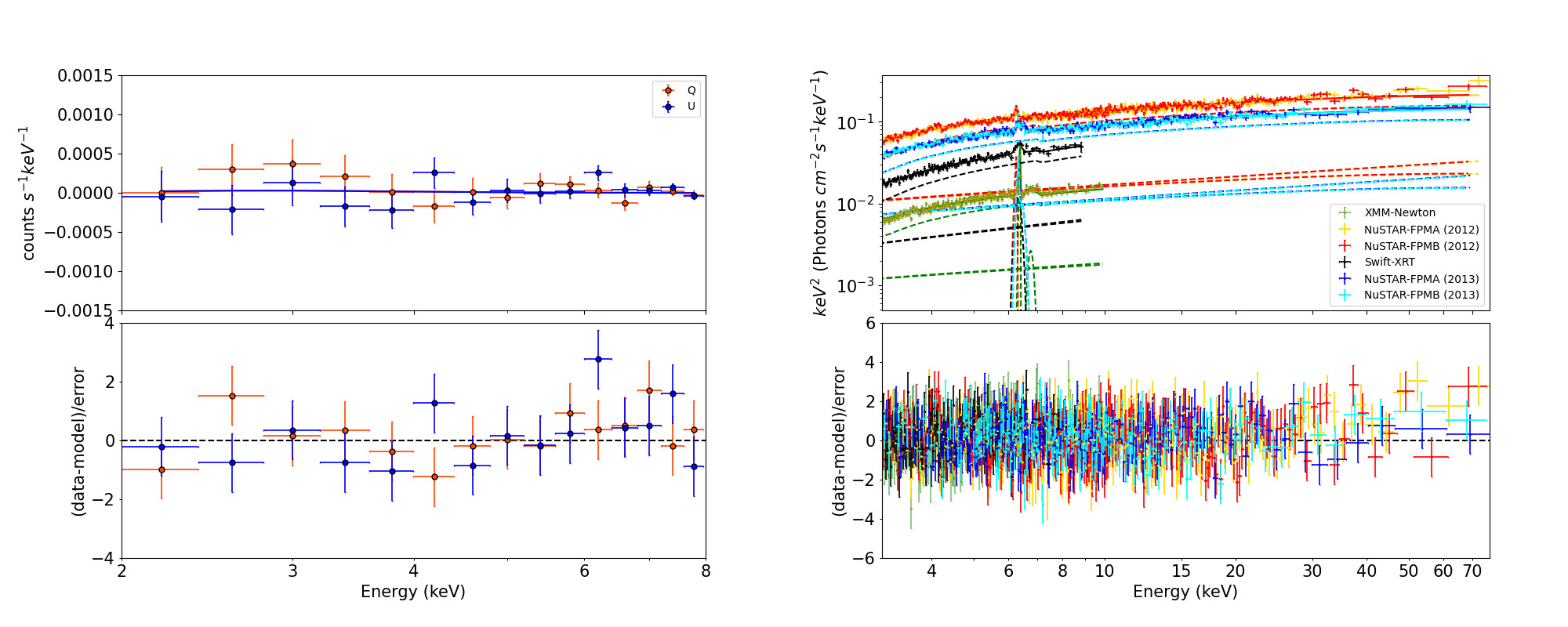}
     }
\caption{Left panel: { IXPE Q, U} stokes best-fit data with model: $constant*phabs*zphabs*(zpo+zgauss)$ and the residuals. Right panel: { NuSTAR, XMM-Newton and Swift-XRT} unfolded best-fitted spectra with model: $constant \times phabs \times (zphabs \times {zcutoff} + pexrav + zgauss)$ (top panel) and the residuals (bottom panel).} 
\label{figure-1}
\end{figure*}

Following \cite{2007ApJ...671.1345E}, we modeled the { XMM-Newton} Fe-K$\alpha$ (6.42$\pm$0.01 keV) and Fe-K$\beta$ (7.01$\pm$0.04 keV) lines with two different $zgauss$ components with their width frozen to 10 eV. We also added a neutral Fe K edge to the model. The { Swift-XRT} observations were fitted together with a Fe-K$\alpha$ line component. From the joint spectral analysis, we derived $\rm{N_{H}} = (5.86^{+0.62}_{-0.61}) \times 10^{22}$ atoms $\rm{cm^{-2}}$, $\Gamma = 1.71^{+0.01}_{-0.02}$, and $\rm{E_{cut}}$ = 280$^{+65}_{-54}$ keV. The best fit spectra with residuals are plotted in Fig. \ref{figure-1}. Our measurements of $\rm{N_{H}}$ and $\Gamma$ are consistent with previous results obtained from {Suzaku} \citep{2014ApJ...786..126R, 2016ApJS..225...14K, 2020ApJ...897....2T} as well as {NuSTAR} observations \citep{2015MNRAS.447..160M, 2020ApJ...905...41B, 2023MNRAS.524.4670J}. While \cite{2020ApJ...905...41B} and \cite{2023MNRAS.524.4670J} reported $\rm{E_{cut}}$ = 300$^{+50}_{-30}$ keV and 218$^{+35}_{-47}$ keV, respectively, \cite{2015MNRAS.447..160M} provided a lower limit of $\rm{E_{cut}} > 210$ keV.

Furthermore, fitting the Comptonization model, $ constant*phabs*zphabs*(xillverCP)$ to the { NuSTAR}, { XMM-Newton} and { XRT} data allowed us to put a tighter constraint on the inclination angle compared to the { NuSTAR}-only fit. The best fit model parameters for the joint fit are given in Table \ref{table-5}.

\begin{table}
        \centering
        \caption{Spectral analysis results from the joint analysis of { NuSTAR}, { XMM-Newton} and { XRT} observations.}
        \label{table-5}
\begin{tabular}{ccc}
\hline
Model component & Parameter & Best-Fit \\
        \hline
$zphabs$ & $\rm{N_{H.los}}$ ($\times$ $10^{22}$) $\rm{cm^{-2}}$ &  5.86$^{+0.62}_{-0.61}$ \\
\hline
$pexrav$ & $\Gamma$ & 1.71$^{+0.01}_{-0.02}$  \\
& $\rm{E_{cut}}$ (keV) &  280$^{+65}_{-54}$ \\
& norm ($\times$ $10^{-2}$) & 3.82$^{+0.10}_{-0.21}$ \\
& $\chi^2/dof$ &  3805/3727 \\
\hline
$xillverCP$ & $\Gamma$ & 1.78$^{+0.01}_{-0.01}$  \\
& $\rm{kT_{e}}$ & $>$54 \\
& $R$ & 0.40$^{+0.07}_{-0.06}$  \\
& norm ($\times$ $10^{-3}$) & 1.77$^{+0.01}_{-0.01}$ \\
& $\theta$ ($^{\circ}$) & 86.0$^{+0.5}_{-0.6}$ \\
& $\chi^2/dof$ &  3757/3736 \\
\hline
\multicolumn{3}{c}{{ XMM-Newton}} \\ 
$zgauss$ & $Line_{E1}$ (keV) & 6.42$^{+0.01}_{-0.01}$  \\
& $norm_{E1}$ ($\times$ $10^{-5}$) & 4.55$^{+0.40}_{-0.40}$ \\
& $Line_{E2}$ (keV) &  7.01$^{+0.04}_{-0.04}$  \\
& $norm_{E2}$ ($\times$ $10^{-5}$) & 1.07$^{+0.34}_{-0.34}$  \\
\hline
\multicolumn{3}{c}{{ NuSTAR} (2012)} \\ 
$zgauss$ & $Line_{E1}$ (keV) & 6.27$^{+0.08}_{-0.08}$  \\
& $norm_{E1}$ ($\times$ $10^{-5}$) & 8.90$^{+0.34}_{-0.34}$ \\
\hline
\multicolumn{3}{c}{{ NuSTAR} (2013)} \\ 
$zgauss$ & $Line_{E1}$ (keV) & 6.41$^{+0.06}_{-0.06}$  \\
& $norm_{E1}$ ($\times$ $10^{-5}$) & 2.23$^{+0.50}_{-0.50}$ \\
\hline
\multicolumn{3}{c}{{ Swift-XRT}} \\ 
$zgauss$ & $Line_{E1}$ (keV) & 6.39$^{+0.07}_{-0.06}$  \\
& $\sigma$ & $<$0.17 \\
& $norm_{E1}$ ($\times$ $10^{-5}$) & 3.43$^{+0.16}_{-0.15}$ \\
\hline
\end{tabular}
\end{table}

\section{Discussion and summary}
Till now, {IXPE} polarimetric observations have been utilized to investigate the morphology of X-ray coronae in three unobscured radio-quiet Seyfert galaxies: MCG-05-23-16 \citep{2022MNRAS.516.5907M, 10.1093/mnras/stad2627}, NGC 4151 \citep{2023MNRAS.523.4468G, 2024A&A...691A..29G}, and IC 4329A \citep{2023JApA...44...87P, 2023MNRAS.525.5437I}. From the first {IXPE} observation of MCG-05-23-16, \cite{2022MNRAS.516.5907M} reported an lower limit on $\Pi_{X}$ of $\gtrsim$ 4.7\% at the 99\% confidence level. After combining the first observation with the second,  $\Pi_{X}$ of the primary continuum was found to be $\gtrsim$ 3.2\% at the 99\% confidence level \citep{10.1093/mnras/stad2627}. Both observations suggested a hint of alignment between the polarization angle and the accretion disc spin axis, disfavoring a spherical geometry. For IC 4329A, a 3$\sigma$ upper limit on $\Pi_{X}$ of 6.2\% was measured. In NGC 4151, a clear detection was reported with $\Pi_{X} = 4.9 \pm 1.1\%$. In both IC 4329A and NGC 4151, $\Psi_{X}$ was found to be parallel to the disc axis, disfavoring the spherical lamppost geometry.

In this work, we conducted polarimetric and spectro-polarimetric analyses of NGC 2110 using the first { IXPE} observation. A model-independent analysis of its polarization properties, performed using polarization cubes ($ PCUBE$) across multiple energy bands, is presented in Table \ref{table-2}. In each energy band, the measured polarization degree remains below the MDP values, preventing us from claiming a significant polarization detection for NGC 2110. As a result, the polarization angle remains unconstrained across all energy bands. While Table \ref{table-2} suggests an increase in polarization degree with energy, this trend is likely a consequence of rising upper limits due to the decreasing signal-to-noise ratio at higher energies. Our model-dependent analysis in the 2–8 keV band supports the $ PCUBE$ results, placing an upper limit of 7.6\% on $\Pi_X$ at the 99\% confidence level. Furthermore, our analysis of the 5.66–8 keV { IXPE I, Q, and U} spectra, including contour plots between $\Pi_X$ and $\Psi_X$ (Fig. \ref{figure-2}, right panel), constrains the upper limit of $\Pi_X$ to $<$27\% at the 99\% confidence.

From an analysis of NGC 2110, VLA 3.6 and 20 cm observations, \cite{1999ApJS..120..209N} reported a radio jet position angle (PA) of 9$^{\circ}$. Using VLBA 8.4 GHz radio continuum observations, \cite{2000ApJ...529..816M} measured a North-South radio jet PA of 8.5$^{\circ} \pm 0.8^{\circ}$. However, the upper limits on $\Pi_X$ obtained from both our polarimetric and spectro-polarimetric analyses of the primary emission in the 2$-$8 keV band prevent us from placing a meaningful constraint on $\Psi_X$, making it difficult to draw definitive conclusions about the coronal geometry of NGC 2110. Additionally, the contour plot between $\Pi_X$ and $\Psi_X$ does not allow us to determine whether the radio jet position angle is aligned parallel or perpendicular to $\Psi_X$.

The earlier analysis for MCG-05-23-16 and NGC 4151 \citep{2022MNRAS.516.5907M, 10.1093/mnras/stad2627, 2023MNRAS.523.4468G} shows that $\Psi_X$ being parallel to the accretion disk disfavors a spherical or cone-shaped corona considering the systems inclinations $\gtrsim$ 40$^{\circ}$–50$^{\circ}$, which are typical for Seyfert galaxies. A polarization angle parallel to the accretion disk axis indicates that a flat configuration of the emitting matter is more likely to produce such a scenario. This is supported by simulations of polarization signals for various coronal geometries, including slab, cone, and spherical models \citep{10.1093/mnras/stab3745}. Using the  $ MONK$ \citep{Zhang_2019} code, \cite{10.1093/mnras/stab3745} demonstrated in their simulations that $\Psi_X$ for slab and spherical/conical coronas is not strongly dependent on energy or system inclination. There is always a 90$^{\circ}$ difference in $\Psi_X$ between the slab and cone or spherical corona. While for the slab geometry, $\Psi_X$ is parallel to the disk axis, for conical/spherical ones it is perpendicular to the accretion disk axis. The simulation also revealed that $\Pi_X$ is influenced by coronal parameters ($\Gamma$, $\rm{kT_{e}}$, $\tau$), corona size/black-hole spin, and system inclination. At low inclination, it is challenging to distinguish between coronal geometries when $\Pi_X \sim 0$–2\%. However, for high-inclination systems, distinct differences in $\Pi_X$ emerge among geometries (see Fig. 2 of \citealt{10.1093/mnras/stab3745}). At high inclinations, a slab corona produces polarization degrees up to 12\%. Due to the symmetrical nature of a spherical corona, the expected polarization degree is significantly lower ($\sim 1$–3\%) compared to a slab geometry \citep{1996ApJ...470..249P, 2018A&A...619A.105T}. For a conical corona, the expected polarization lies between that of the slab and spherical geometries. Recently, \cite{10.1093/mnras/stad2627} used the $ MONK$ code to simulate polarization signals for a wedge-shaped corona in MCG-05-23-16. They found that the wedge-shaped geometry is similar to the slab, producing $\Psi_X$ parallel to the accretion disk axis. Moreover, the wedge-shaped configuration resolves the issue of slab geometry producing only super-soft ($\Gamma \geq 2.0$) X-ray spectra \citep{1995ApJ...449L..13S, 2007A&ARv..15....1D, 2018A&A...614A..79P}.

Our broad-band spectral analysis of the { NuSTAR} observations of NGC 2110 using the Comptonization model, $xillverCP$ yielded a photon index, $\Gamma$ of 1.74$\pm$0.01 and a lower limit of the electron temperature, $\rm{kT_{e}}$ of 74.86 keV, implying an optical depth, $\tau$ of $<$ 1.61. Additionally, we derived a lower limit on the inclination angle, finding $i$ $>$ 74$^{\circ}$. The joint analysis of the { NuSTAR}, { XMM-Newton}, and { Swift-XRT} data using the same Comptonization model produced consistent spectral parameters (see Table \ref{table-5}). Comparing our best-fit values of $\rm{kT_e}$ and inclination angle with the simulations from \cite{2022MNRAS.516.5907M} and \cite{10.1093/mnras/stad2627} suggests that, while the high inclination of NGC 2110 makes it a promising candidate for polarization studies to probe its coronal geometry, the current polarization measurements lack the sensitivity to place definitive constraints on the polarization parameters. Contour plots from our XSPEC analysis indicate that the upper limit on $\Pi_{X}$ (at the 99\% confidence in the 2–8 keV band) is approximately 8\% if the polarization is aligned with the radio jet (suggesting a corona elongated along the disk) and around 4\% if perpendicular (consistent with a more spherical corona). Given these constraints, the existing data do not allow us to draw definitive conclusions about the coronal structure of NGC 2110.

\begin{acknowledgement}
We thank the referee for their valuable suggestions, which have helped improve the clarity of the paper.  The { IXPE} data used in this paper are publicly available in the HEASARC database:  
\url{https://heasarc.gsfc.nasa.gov/docs/ixpe/archive/}. We thank the { NuSTAR} Operations, Software, and Calibration teams for their support in the execution and analysis of these observations. This research has made use of archival data from the { XMM-Newton} and { NuSTAR} observatories, accessed through the High Energy Astrophysics Science Archive Research Center (HEASARC) online service, provided by the NASA Goddard Space Flight Center.     
\end{acknowledgement}

\bibliographystyle{aa} 
\bibliography{example}

\label{lastpage}
\end{document}